# Complete determination of crystallographic orientation of ReX$_2$ (X=S, Se) by polarized Raman spectroscopy

*Yun Choi,[†] Keunui Kim,[†] Soo Yeon Lim,[†] Jungcheol Kim,[†] Je Myoung Park,[†] Jung Hwa Kim,[‡] Zonghoon Lee,[‡, §] and Hyeonsik Cheong[†,*]*

[†]Department of Physics, Sogang University, Seoul 04107, Korea

[‡]School of Materials Science and Engineering, Ulsan National Institute of Science and Technology, Ulsan 44919, Korea

[§]Center for Multidimensional Carbon Materials, Institute for Basic Science (IBS), Ulsan 44919, Korea

[*]E-mail: hcheong@sogang.ac.kr

**Abstract**

Polarized Raman spectroscopy on few-layer $ReS_2$ and $ReSe_2$ was carried out to determine the crystallographic orientations. Since monolayer $ReX_2$ (X=S or Se) has a distorted trigonal structure with only an inversion center, there is in-plane anisotropy and the two faces of a monolayer crystal are not equivalent. Since many physical properties vary sensitively depending on the crystallographic orientation, it is important to develop a reliable method to determine the crystal axes of $ReX_2$. By comparing the relative polarization dependences of some representative Raman modes measured with three different excitation laser energies with high-resolution scanning transmission electron microscopy, we established a reliable procedure to determine the all three principal directions of few-layer $ReX_2$ including a way to distinguish the two types of faces: a 2.41-eV laser for $ReS_2$ or a 1.96-eV laser for $ReSe_2$ should be chosen as the excitation source of polarized Raman measurements; then the relative directions of the maximum intensity polarization of the Raman modes at 151 and 212 $cm^{-1}$ (124 and 161 $cm^{-1}$) of $ReS_2$ ($ReSe_2$) can be used to determine the face types and the Re-chain direction unambiguously.

**Conceptual insights**

The crystallographic anisotropy of rhenium disulfide and diselenide ($ReS_2$ and $ReSe_2$) gives additional degrees of freedom in manipulating device properties. In addition to having in-plane anisotropy, the two sides of these crystals are not equivalent because of the low symmetry of $C_i$. Since the electrical and optical properties depend sensitively on the anisotropic orientation of the crystal, it is very important to develop a reliable method to determine the crystallographic orientation, including a method to determine the up- and down-side of the crystal. In this work, we developed the first comprehensive and unambiguous method to determine the crystallographic orientation of few-layer ReS$_2$ and ReSe$_2$ by combining polarized Raman spectroscopy with high-resolution electron microscopy. The results are confirmed by flipping a sample over and repeating the measurements on both sides of the same sample. To our surprise, the choice of the excitation laser turns out to be crucial, especially for ReSe$_2$. We also found that the maximum intensity polarization direction of the Raman mode is slightly offset from the principal axis in the case of a monolayer, whereas the two directions match for thicker samples. Such detailed information is crucial in precisely determining the crystal axes for other applications.

**Introduction**

The 2-dimensional (2D) layered transition metal dichalcogenide (TMD) ReX$_2$ (X=S, Se), are attracting much interest for application in optoelectronic devices because ReS$_2$ has a direct bandgap for all thicknesses from monolayer (1L) to bulk unlike other TMDs such as MoX$_2$ and WX$_2$ (X=S, Se) which have a direct bandgap only for the monolayer case.[1–4] The optical bandgap energy of ReS$_2$ varies in the visible range, from 1.47 eV for the bulk to 1.61 eV for 1L, depending on the number of layers,[5–7] making it an attractive candidate for optoelectronic devices. On the other hand, ReSe$_2$ has an indirect (nearly direct) band gap of 1.2 eV in monolayer and 1.3 eV in bulk.[8–11] Unlike group-VI TMDs, extra electrons of ReX$_2$ (group-VII TMDs) make Re-Re bonds resulting in Peierls distortion.[5,12–14] Due to this covalent bonding of the transition metal atoms (Re-Re), ReX$_2$ has a distorted trigonal structure (1T′) in which the principal axes are not normal to each other, i.e., the triclinic structure with the space group $P\bar{1}$ (No. 2).[8,13,15–19] Because of the 1T′ structure, ReX$_2$ has in-plane anisotropy like black phosphorus[20] or WTe$_2$,[21,22] which leads to anisotropic optical and electronic properties[23–30] that can be utilized for polarization-sensitive photodetector[31,32] or orientation-dependent thin film transistors.[26] Furthermore, since 1L ReX$_2$ has only one inversion center in the primitive cell. The two faces of 1L ReX$_2$ are not equivalent. For optoelectronic devices, determining the orientation of the ReX$_2$ layer (up or down) is as important as identification of the in-plane crystallographic orientation. Therefore, an easy method to determine both the in-plane crystallographic orientation and the vertical orientation (up or down) of ReX$_2$ is needed to fabricate devices with orientation-dependent properties.

Raman spectroscopy is widely used in 2D materials research for characterization of several basic properties including the number of layers, strain, doping, etc. Polarized Raman spectroscopy, in particular, has proven to be a powerful tool to determine to determine the crystal structures of 1D or 2D nanomaterials such as nanowires or black phosphorus crystals.[19–22,33–38] Some polarized Raman spectroscopy studies have been carried out to determine the orientation of ReX$_2$.[37,38] The in-plane crystal axes of bulk and few-layer ReX$_2$ could be determined by comparing the polarization directions of some Raman modes with the transmission electron microscopy (TEM) measurements.[19,37] However, as we shall show in this work, the choice of the excitation laser energy is crucial because the polarization directions of the Raman modes depend on the excitation laser energy sensitively. In addition, the two vertical orientations were identified in bulk crystals of ReX$_2$, but similar work on few-layer ReX$_2$ has not been reported. Here, we report on a comprehensive set of polarized Raman measurements on few-layer ReX$_2$ using several excitation energies in combination with high-resolution scanning transmission electron microscopy (HR-STEM) to determine the in-plane axis and the vertical orientation simultaneously. We show that the 633-nm (1.96-eV) excitation for ReSe$_2$ and the 514.5-nm (2.41-eV) excitation for ReS$_2$, respectively, are most suitable for reliable determination of the crystallographic orientations.

**Results and Discussion**

Fig. 1a shows the crystal structure of 1L ReX$_2$, which comprises a rhenium layer sandwiched between two sulfur layers. Because 1L ReX$_2$ belongs to the point group C$_i$ which has inversion symmetry only, it has two types of top-view as shown in Fig. 1a.[19,39] The Re chain direction is taken as the *b*-axis and the other in-plane principal axis is taken as the *a*-axis. Then the (bulk) *c*-axis has opposite directions in the two types. We identify the two orientations in terms of the relative direction of the *c*-axis of the bulk crystal. The orientation in which the bulk *c*-axis is coming out of the plane will be called as the '*c*-up' type and the other as the '$\bar{c}$-up' type. We should note that because the *c*-axis is not orthogonal to *a*- or *b*-axis due to the triclinic structure, the *c*-axis is slightly tilted from the face normal. Some of the cleaved edges of exfoliated ReX$_2$ samples are found to be approximately 60° or 120° with respect to each other as shown in Fig. 1b and c, and are usually assumed to be *a*- or *b*-axis.[19,38,40] However, as we will see below, assigning the crystallographic direction from the edge directions is often inaccurate.

The unit cell of 1L ReX$_2$ has 12 atoms, and the irreducible representation can be written as $\Gamma = 18(A_g + A_u)$, among which $A_g$ modes are Raman active.[8,18,41] The 18 Raman-active $A_g$ type modes of ReS$_2$ are observed in the range of 130 to 440 cm$^{-1}$ and those of ReSe$_2$ in the range of 100 to 300 cm$^{-1}$, as shown in Fig. 1e and g, respectively.[18,25,39] (see Supplementary Information Fig. S1 for Raman spectra measured with different excitation energies) These modes are labelled in the order of the frequency as mode 1 through mode 18 as indicated and the peak positions are tabulated in Table S1. Raman spectra measured in several polarization directions are shown for comparison; the relative angle between the *b*-axis (Re-chain direction) and the polarization directions of the excitation and scattered photons (parallel polarization configuration) is varied. Some of the modes show prominent polarization dependences. (see Supplementary Information Fig. S2 for the polarization dependence of all 18 modes of 1L ReX$_2$) In addition, the inter-layer shear and breathing modes in the ultralow-frequency range below 40 cm$^{-1}$ measured in the parallel polarization configuration are observed for 2L or thicker samples as shown in Fig. 1d and f. These modes are the most reliable indicators of the number of layers.[42,43] Although there have been reports in which the polarization dependences of these modes were examined for the determination of the crystal structures. However, we concentrate on the higher-frequency intralayer vibration modes that are common to all thicknesses. We also observed the Brillouin-scattering peak from the Si substrate (~5cm$^{-1}$) which is marked by '*' in Fig. 1d and f.[44–46] Among the 18 first-order intra-layer Raman modes, the lower frequency modes tend to be stronger and have more straightforward polarization dependences. We focused on the 5 lowest frequency intra-layer modes for the polarization analysis. (see Supplementary Information Fig. S3 and 4 for the polarization dependence of 5 lowest frequency modes of few-layer ReX$_2$) For ReS$_2$, modes 1, 3, and 5 are relatively strong and have simple polarization dependences. Also, because these modes do not overlap with other modes, it is easier to analyze the polarization dependence from the spectra. We compared modes 3 and 5 for the analysis. For ReSe$_2$, modes 1, 4, and 5 are spectrally isolated and have relatively simple polarization dependences. Since mode 1 is rather weak, we chose modes 4 and 5 for the analysis. Although all the modes are $A_g$ type, the polarization dependences are dramatically different.[19,42] The polarization dependence can be analysed in terms of the Raman tensors of the $A_g$ modes of the $C_i$ point group:[37,42]

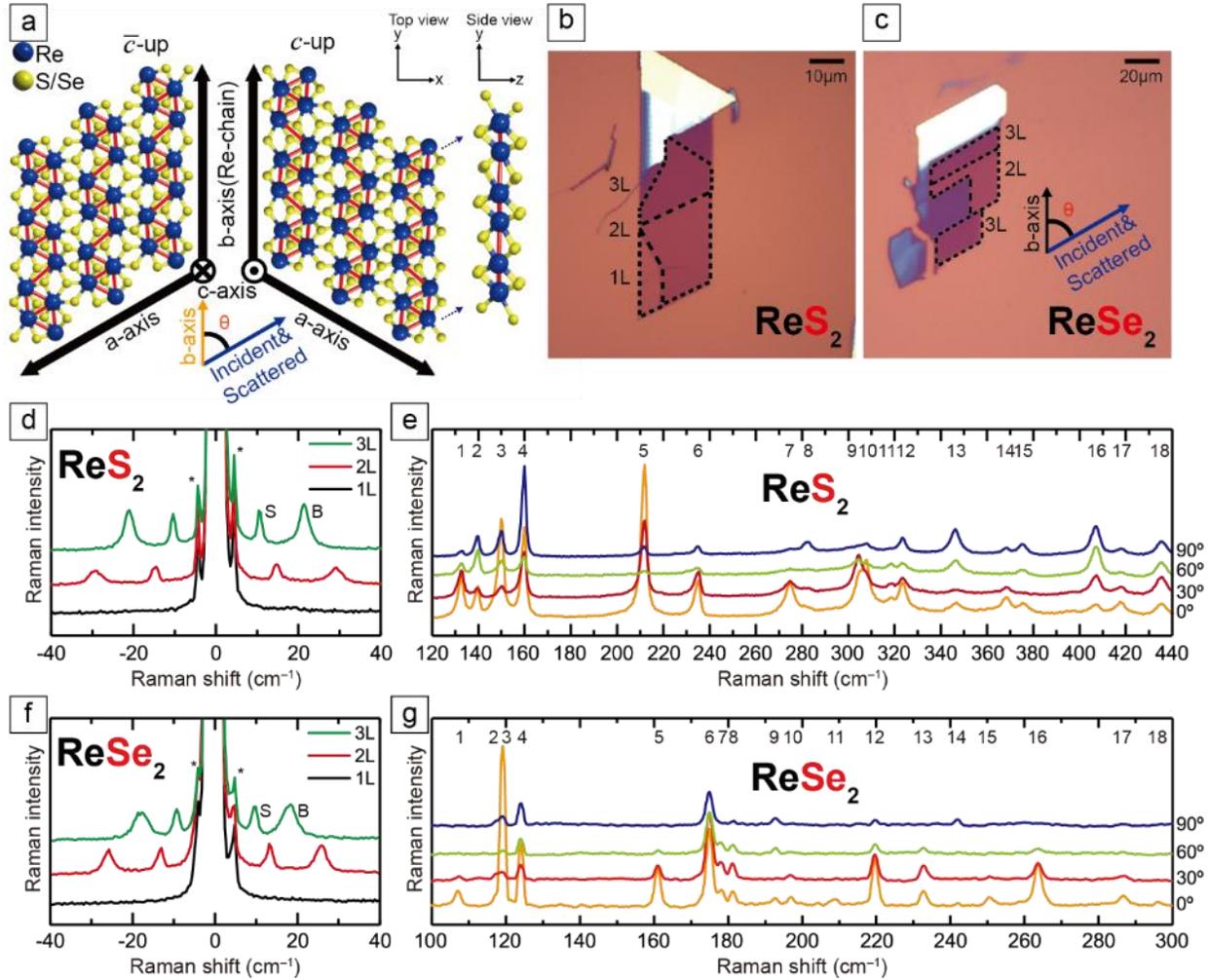

**Fig. 1** (a) Top- and side-view of crystal structure of 1L ReX$_2$. The principal axes are indicated. The 'c-up' type has the c-axis coming out of the plane and the '$\bar{c}$-up' type into the plane. (b, c) Optical images of ReX$_2$ samples with 1L, 2L, 3L areas indicated. (d, f) Low-frequency Raman spectra of 1 to 3L ReX$_2$ measured by using a 2.33-eV laser. (e, g) Raman spectra of 1L ReX$_2$ for various polarization directions with respect to the b-axis. (2.41-eV laser for ReS$_2$, 1.96-eV laser for ReSe$_2$)

$$R(A_g) = \begin{pmatrix} a & d & e \\ d & b & f \\ e & f & c \end{pmatrix}. \quad (1)$$

In anisotropic 2D materials, the polarization-dependent absorption and birefringence can be taken into account by expressing the Raman tensor elements in complex forms.[20,21,42,47–53] The Raman tensors can be written as

$$R(A_g) = \begin{pmatrix} |a|e^{i\phi_a} & |d|e^{i\phi_d} & |e|e^{i\phi_e} \\ |d|e^{i\phi_d} & |b|e^{i\phi_b} & |f|e^{i\phi_f} \\ |e|e^{i\phi_e} & |f|e^{i\phi_f} & |c|e^{i\phi_c} \end{pmatrix}. \quad (2)$$

The Raman intensity for excitation and detection in the z-direction is,

$$I(A_g) \propto |\hat{e}_s \cdot R(A_g) \cdot \hat{e}_i|^2. \quad (3)$$

where $\hat{e}_i$ and $\hat{e}_s$ are the polarizations of the incident and scattered photons, respectively. In the backscattering geometry with the parallel polarization configuration, $\hat{e}_i = \hat{e}_s = (\cos\theta, \sin\theta, 0)$, and the Raman intensity is

$$\begin{aligned} I(A_g) \propto & |a|^2 \cos^4\theta + |d|^2 \sin^4\theta + 4|b|^2 \sin^2\theta \cos^2\theta \\ & + 2|a||d|\cos\phi_{ad} \sin^2\theta \cos^2\theta \\ & + 4|a||b|\cos\phi_{ab} \sin\theta \cos^3\theta \\ & + 4|d||b|\cos\phi_{db} \sin^3\theta \cos\theta \end{aligned}, \quad (4)$$

where $\phi_{ab} = \phi_a - \phi_b$, $\phi_{ad} = \phi_a - \phi_d$ and $\phi_{db} = \phi_d - \phi_b$.

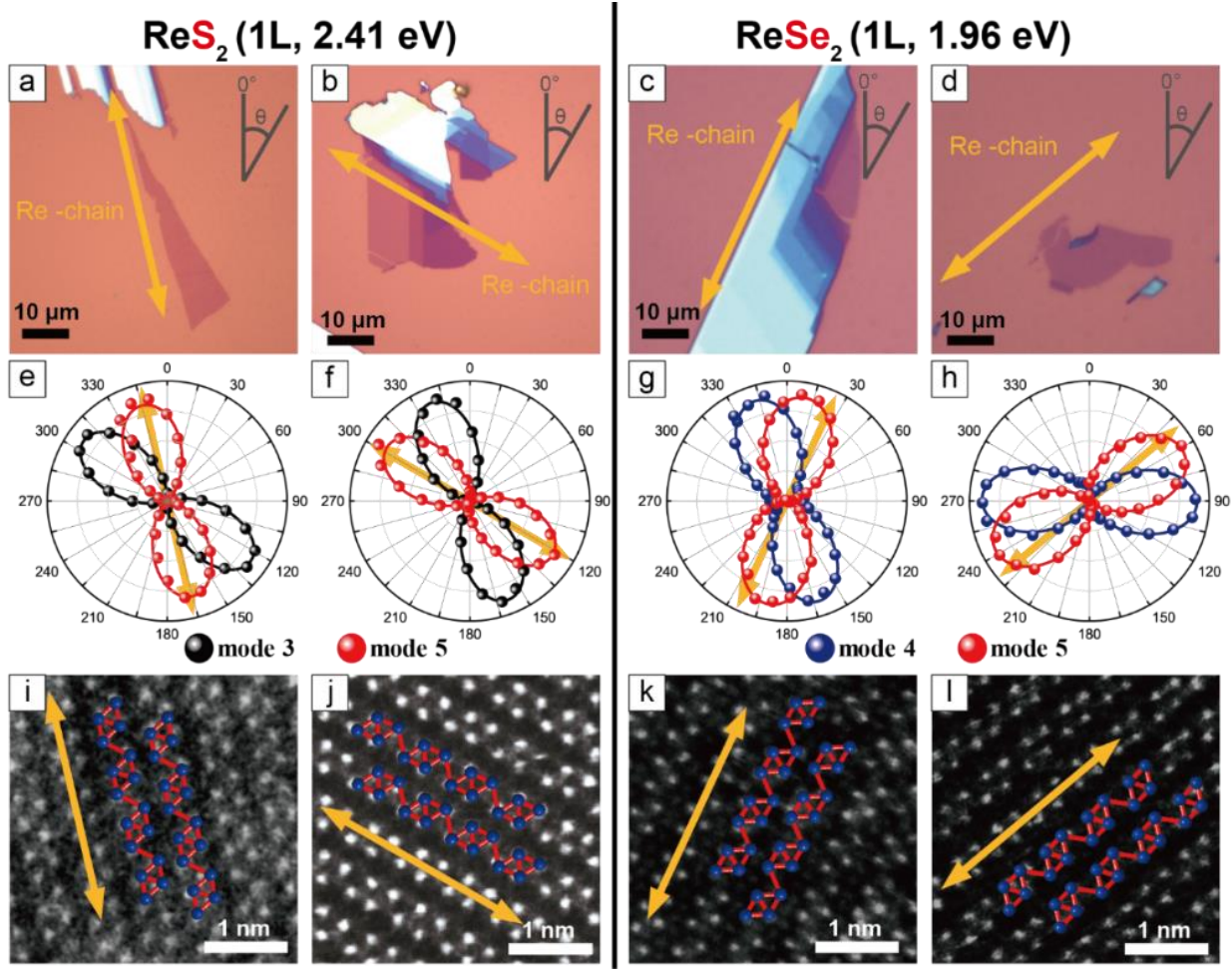

**Fig. 2** (a-d) Optical images of 1L ReS$_2$ and ReSe$_2$ samples. (e-h) polarization dependences of the intensities of Raman modes 3 and 5 at 151 and 212 cm$^{-1}$ for ReS$_2$ (modes 4 and 5 at 124 and 160 cm$^{-1}$ for ReSe$_2$), respectively, measured by using a 2.41-eV laser (1.96-eV laser for ReSe$_2$). (i-l) HR-STEM images of samples. The Re-chain directions are indicated by the arrows. The orientations are (i, k) ′$c$-up′ and (j, l) ′$\bar{c}$-up′.

Fig. 2a-d are optical microscope images of some representative 1L ReX$_2$ samples. The corresponding polar plots (Fig. 2e-h) show the polarization dependences of the intensities of the two chosen modes (3 and 5 for ReS$_2$ and 4 and 5 for ReSe$_2$) measured in the parallel polarization

configuration. There are two types of the relative orientations of the two modes: in Fig. 2e (2g), the maximum direction of mode 3 (4) is slightly rotated in the counter-clockwise direction with respect to that of mode 5, whereas in Fig. 2f (2h) the relative rotation is clockwise. In order to correlate these types with the crystallographic orientations, we determined the Re-chain direction ($b$-axis) from HR-STEM measurements shown in Fig. 2i-l. The HR-STEM images show that the two different types of the relative orientations of the two Raman modes correspond to the two different in-plane structures, i.e., vertical orientations: by comparing with the crystal structure of Fig. 1a, Fig. 2i and k corresponds to the '$c$-up' type and Fig. 2j and l the '$\bar{c}$-up' type. The Re-chain directions thus determined are also shown in the optical images and the polar plots of the Raman intensities by yellow arrows. It is obvious that the Re-chain direction does not necessarily corresponds with the edge direction of a sample. We also note that the maximum intensity directions of the Raman modes do not correspond to the crystallographic orientation exactly. In order to confirm that the two-types of samples correspond to the two vertical orientations, we measured the two sides of an identical sample by placing the sample on a transparent quartz substrate.

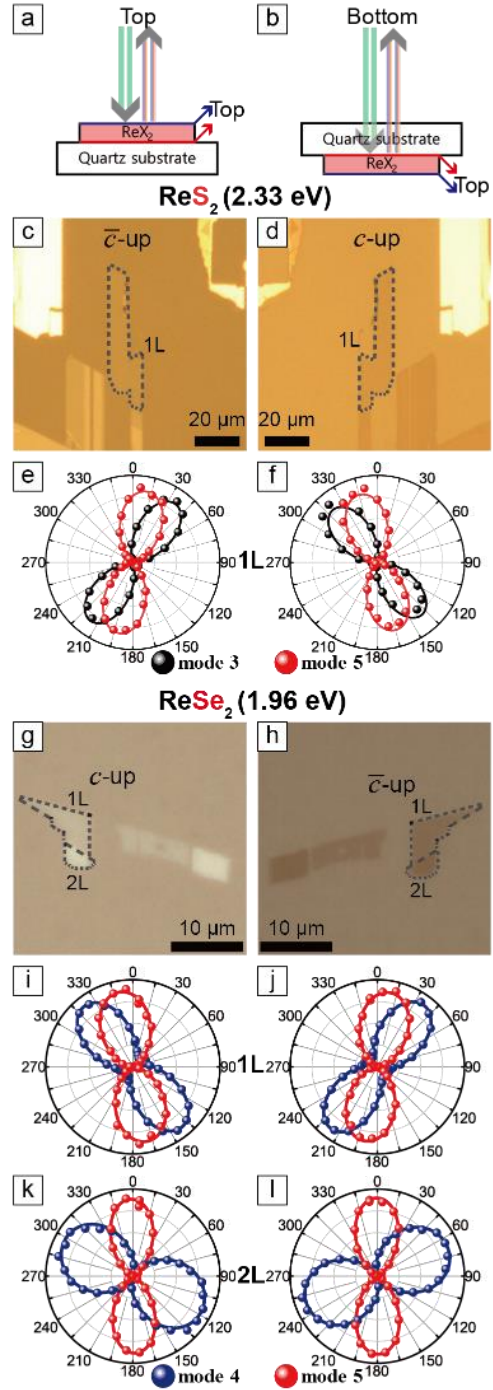

**Fig. 3** (a, b) beam path scheme (c, d, g, h) Optical images 1L ReX$_2$ sample on quartz substrate. (e, f) Polarization dependences of mode 3 and 5 of 1L ReS$_2$ corresponding to c and d, respectively. (i-l) Polarization dependences of mode 4 and 5 for 1L and 2L ReSe$_2$ corresponding to g (i and k) and h (j and l).

Fig. 3c and d are optical images of an 1L ReS$_2$ sample taken by flipping the sample over as shown in Fig. 3a and b. The relative polarization dependences of modes 3 and 5, shown in Fig. 3e and f, correspond to those of the '$\bar{c}$-up' and the '$c$-up' cases in Fig. 2. Similar measurements on 1L and 2L ReSe$_2$ also yield consistent results as shown in Fig. 3. These experiments unambiguously confirm that the relative rotation of the maximum intensity polarization directions of the two modes is a reliable indicator of the vertical orientation.

In order to correlate the polarization dependence of Raman peaks with the in-plane crystallographic axes, we carried out polarized Raman measurements with three excitation energies each for ReS$_2$ and ReSe$_2$. Fig. 4a shows the polarization dependences of modes 3 and 5 of 1L to 3L ReS$_2$ with the fitting curves to Eq. (4), for three excitation energies of 1.96, 2.33, and 2.41 eV. The Re-chain direction ($b$-axis) determined from HR-STEM measurements is set to 0°. In order to ensure that the crystal axes in samples with different thicknesses are properly aligned, we measured a sample with all three thickness in a single contiguous flake. The polarization dependences for the '$c$-up' and '$\bar{c}$-up' cases are mirror images of each other with respect to the $b$-axis (0°). Both modes show 'figure 8' type polarization dependences for all excitation energies and thicknesses. The maximum intensity direction of mode 5 is approximately along the $b$-axis. For 1L, it is rotated with respect to the $b$-axis direction by 5°, –10°, and –3° for the '$c$-up' case for excitation energies of 1.96, 2.33, and 2.41 eV, respectively. For 2L and 3L, the offset angle is zero within the experimental uncertainties. Although all three excitation energies are adequate for determination of the crystal axes, we propose that 2.41 eV is most suitable because the offset between the $b$-axis and the maximum intensity direction of mode 5 is smallest for the 1L case. It is much more complicated for ReSe$_2$. Fig. 4b show that the polarization dependences of modes 4 and 5 change dramatically with the excitation energy and thickness. The two modes maintain the 'figure 8' polarization dependences only for the 1.96-eV excitation energy. Here, the offset between the *b*-axis and the maximum intensity direction of mode 5 is –10° for the '*c*-up' type of 1L and is zero for thicker layers. Fig. 4 Polarization dependences of Raman modes (ReS$_2$: 3 and 5, ReSe$_2$: 4 and 5) tabulated with excitation source energies and thickness of ReX$_2$. The correlation between the maximum intensity polarization directions of the modes and the crystal axes can be qualitatively understood in terms of the normal mode vibrations. For both ReS$_2$ and ReSe$_2$, mode 5 involves vibrations of the Re atoms along the *b*-axis.[5,25] Therefore, it is reasonable that the maximum polarization direction (almost) matches that of the *b*-axis. On the other hand, mode 3 of ReS$_2$ involves vibrations of Re and S atoms along the *a*-axis[5] and the maximum intensity direction is approximately in that direction. Since the relative angle between the *a*- and *b*-axes are opposite for the two vertical orientations, it can be used to identify the vertical orientations. For ReSe$_2$, mode 4 involves vibrations of Re and Se atoms in different directions, with several atoms in the unit cell vibrating in directions somewhat tilted from the *a*-axis,[25] which we presume, is why mode 4 is approximately aligned along the *a*-axis but the polarization dependence varies significantly with the excitation energy.

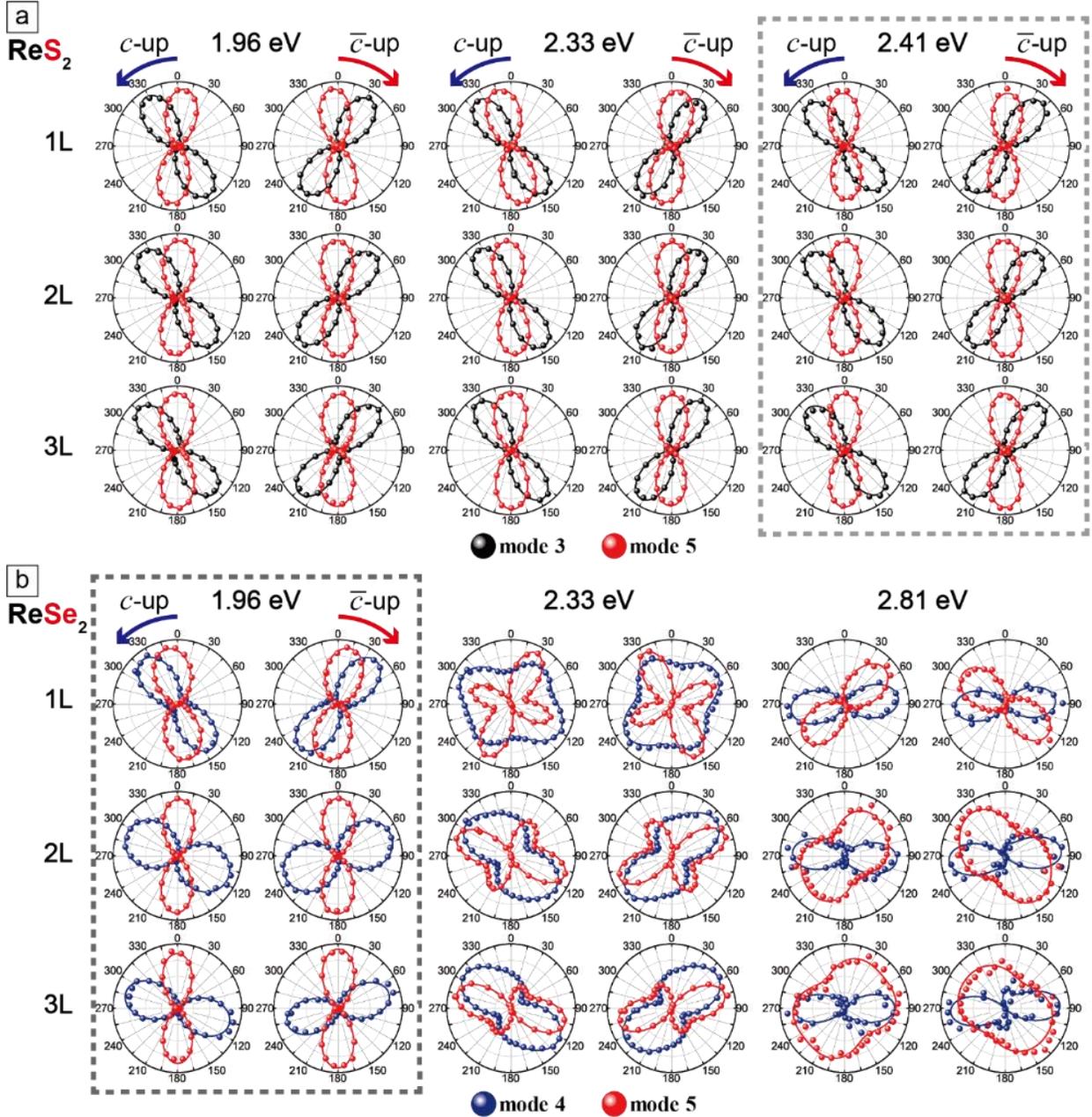

**Fig. 4** Polarization dependences of Raman modes (ReS$_2$: 3 and 5, ReSe$_2$: 4 and 5) tabulated with excitation source energies and thickness of ReX$_2$. The solid curves are fits to Eq. (4).

The slight offset between the *b*-axis and the maximum intensity direction of mode 5 in the 1L samples of both ReS$_2$ and ReSe$_2$ is intriguing. We have measured multiple samples to confirm that this is reproducible. Some possible causes of the offset are residual strain in the sample and the influence of the SiO$_2$/Si substrate. In order to check this possibility, we compared the polarization dependence of mode 5 of a free-standing ReS$_2$ sample with that from a sample on a substrate and found no difference (see Supplementary Information Fig. S5). We also examined the effect of a uniaxial strain by exfoliating a ReS$_2$ sample on a flexible substrate and applying a tensile strain, but the polarization dependences of modes 3 and 5 were not affected (see Supplementary Information Fig. S6). We also fabricated a ReS$_2$ sample with contiguous 1 to 3L regions on hBN and found that the relative rotation of the mode 5 in 1L ReS$_2$ with respect to those in 2L and 3L regions was identical to the case of samples on SiO$_2$/Si (see Supplementary Information Fig. S7). Therefore, we conclude that the slight offset of the maximum polarization direction of mode 5 with respect to the *b*-axis direction in the 1L case is an intrinsic property. In the literature, there are conflicting reports on the alignment of the maximum intensity direction of mode 5 in ReX$_2$ with the *b*-axis. Our results show beyond experimental uncertainty that the two are aligned in 2L and 3L but there is a clear offset in the 1L case. We suspect that the details of the band structure would be responsible for the peculiar polarization for the 1L case. For example, Gehlmann *et al*. examined the band structure of ReS$_2$ and found that the top of the valence band maximum at the Γ point is very flat unlike other thicknesses.[7]

**Experimental**

Few-layer ReX$_2$ samples were prepared by mechanical exfoliation from ReX$_2$ flakes (HQ Graphene) on Si substrates with a 280-nm SiO$_2$ layer and on 0.5 cm quartz substrates. It has been reported that ReS$_2$ can exist in two different polytypes: isotropic-like (IS) and anisotropic-like (AI).[54] Since IS types are much more common, we selected only IS-type samples for our study.

For ReSe$_2$, we found only one type of samples judged from their Raman spectra. In the case of ReSe$_2$, the samples were kept in vacuum to avoid degradation because thin ReSe$_2$ sample seems to degrade after measured with the 2.81-eV laser although the sample remained stable in ambient conditions without laser exposure. We identified the number of layers by comparing the optical contrast and Raman measurements.

The Raman measurements were performed by using a home-built confocal micro-Raman system using the excitation sources of the 2.81-eV (441.6-nm) line of a He-Cd laser, the 2.41-eV (514.5-nm) line of an Ar ion laser, the 2.33-eV (532-nm) line of a diode-pumped-solid-state laser, and the 1.96-eV (632.8-nm) line of a He-Ne laser. ReSe$_2$ samples were kept in an optical vacuum chamber to avoid laser-induced damage to the sample, whereas ReS$_2$ samples were measured in ambient conditions. A 50× objective lens (0.8 NA) and a 40× objective lens (0.6 NA) were used to focus the laser beam onto the sample and to collect the scattered light (backscattering geometry). The Raman signal was dispersed with a HORIBA iHR550 spectrometer (2400 grooves/mm) and detected with a liquid-nitrogen-cooled back-illuminated charge-coupled-device (CCD) detector. The laser power was kept below 0.1 mW. Three volume holographic notch filters (OptiGrate) were used to observe the low frequency region (< 100 cm$^{-1}$). An achromatic half-wave plate was used to rotate the polarization of the linearly polarized laser beam to desired direction. All measurements were conducted in the parallel-polarization configuration, where the analyzer angle was set such that photons with the polarization parallel to the incident polarization pass through. Another achromatic half-wave plate was placed in front of the spectrometer to keep the polarization direction of the signal entering the spectrometer constant with respect to the groove direction of the grating.

The TEM measurements were performed after polarized Raman measurements using an aberration-corrected FEI Titan cube G2 60-300 with a monochromator, operated at 80 kV. Scanning TEM (STEM) analysis were applied for the analysis of definitive orientation of the diamond shaped rhenium chains. For TEM measurements, exfoliated $ReX_2$ flakes on the SiO2/Si substrate were transferred onto a TEM grid using the wet direct transfer method without poly(methyl methacrylate) (PMMA).[55] The orientation of the sample was carefully aligned before and after the transfer in order to match with the orientation of the sample in the polarized Raman measurements.

**Conclusion**

Based on our results, we suggest the following procedure for determining the vertical orientation and the in-plane Re-chain direction of $ReX_2$. First, a 2.41-eV laser for $ReS_2$ or a 1.96-eV laser for $ReSe_2$ should be chosen as the excitation source of polarized Raman measurements. In the case of $ReS_2$, other excitation energies can be used, but then a different offset value should be used in the case of 1L $ReS_2$. For $ReSe_2$, other laser energies should be avoided. From the polarization dependences of modes 3 and 5 (4 and 5) of $ReS_2$ ($ReSe_2$), the vertical orientation can be determined: if mode 3 (mode 4) is somewhat rotated counter-clockwise with respect to mode 5, the sample corresponds to the 'c-up' type $ReS_2$ ($ReSe_2$), and the '$\bar{c}$-up' type if the rotation is clockwise. Then the *b*-axis corresponds to the maximum intensity direction of mode 5 for 2L or thicker samples. For 1L $ReS_2$ ($ReSe_2$), the *b*-axis is offset by 3° (10°) in the clockwise direction with respect to the maximum intensity direction of mode 5 for the 'c-up' case, and in the opposite direction for the '$\bar{c}$-up' case.

**Conflicts of interest**

The authors declare no competing financial interests.

**Acknowledgements**

This work was supported by the National Research Foundation (NRF) grants funded by the Korean government (MSIT) (NRF-2019R1A2C3006189, No. 2018R1A2A2A05019598, and No. 2017R1A5A1014862: SRC program (vdWMRC)), by a grant (2013M3A6A5073173) from the Center for Advanced Soft Electronics under the Global Frontier Research Program of MSIT, and by IBS-R019-D1.

# SUPPLEMENTARY INFORMATION

# Complete determination of crystallographic orientation of ReX$_2$ (X=S, Se) by polarized Raman spectroscopy

*Yun Choi,[†] Keunui Kim,[†] Soo Yeon Lim,[†] Jungcheol Kim,[†] Je Myoung Park,[†] Jung Hwa Kim,[‡] Zonghoon Lee,[‡, §] and Hyeonsik Cheong[†,\*]*

[†]Department of Physics, Sogang University, Seoul 04107, Korea

[‡]School of Materials Science and Engineering, Ulsan National Institute of Science and Technology, Ulsan 44919, Korea

[§]Center for Multidimensional Carbon Materials, Institute for Basic Science (IBS), Ulsan 44919, Korea

[\*]**E-mail: hcheong@sogang.ac.kr**

**Contents:**

● **Figure S1.** Raman spectra of 1L ReS$_2$ and ReSe$_2$ measured with three excitation energies.

● **Table S1.** Measured Raman peak positions of 18 vibrational modes of 1L ReS$_2$ and ReSe$_2$.

● **Figure S2.** Polarization dependences of 18 Raman modes of '*c*-up type' monolayer ReS$_2$ and ReSe$_2$

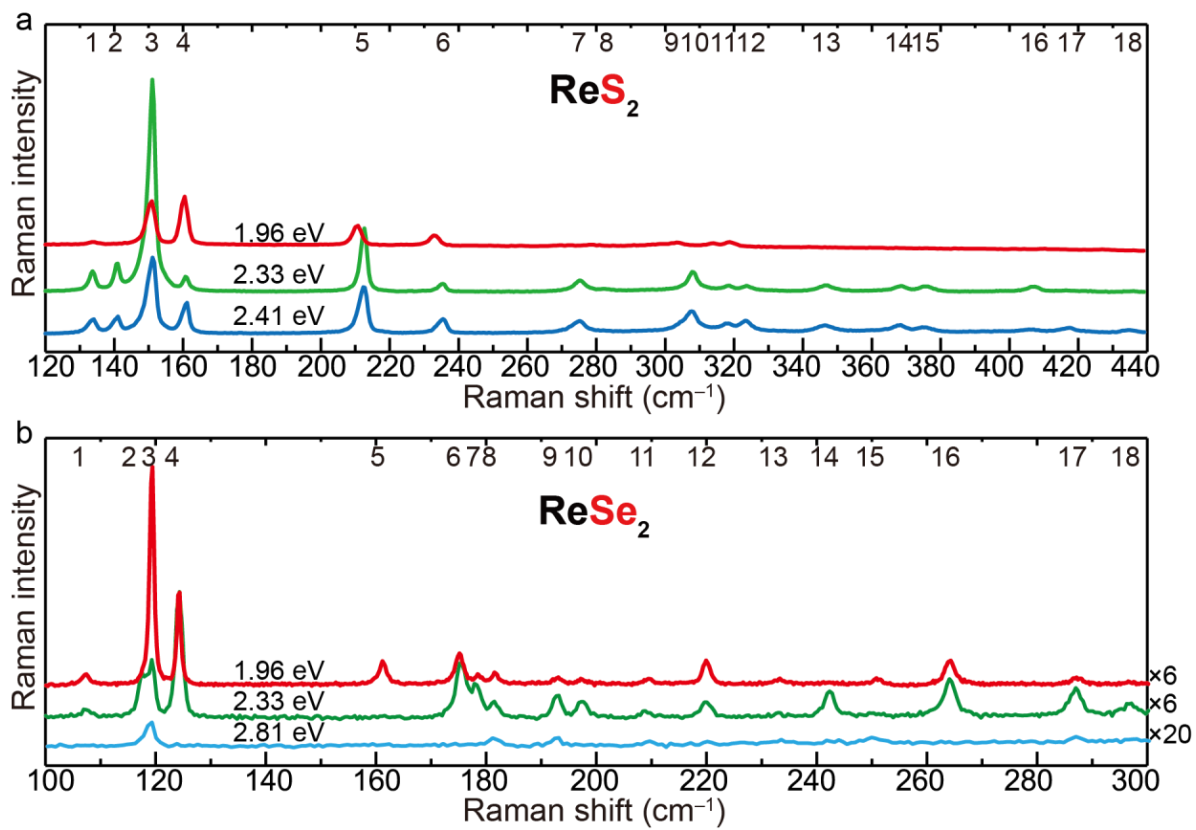

**Fig. S1** Raman spectra of 1L ReS$_2$ and ReSe$_2$ measured with three excitation energies in the parallel-polarization configuration. The polarization angle of 20° with respect to the b-axis was chosen to show all 18 Raman modes: (a) ReS$_2$ and (b) ReSe$_2$.

| mode | ReS$_2$ peak position (cm$^{-1}$) | ReSe$_2$ peak position (cm$^{-1}$) |
| --- | --- | --- |
| 1 | 133 | 108 |
| 2 | 141 | 118 |
| 3 | 151 | 119 |
| 4 | 160 | 124 |
| 5 | 212 | 161 |
| 6 | 236 | 175 |
| 7 | 275 | 178 |
| 8 | 283 | 181 |
| 9 | 305 | 193 |
| 10 | 308 | 197 |
| 11 | 319 | 209 |
| 12 | 324 | 220 |
| 13 | 347 | 233 |
| 14 | 369 | 242 |
| 15 | 376 | 250 |
| 16 | 408 | 264 |
| 17 | 418 | 287 |
| 18 | 436 | 296 |

**Table S1** Measured Raman peak positions of 18 vibrational modes of 1L ReS$_2$ and ReSe$_2$.

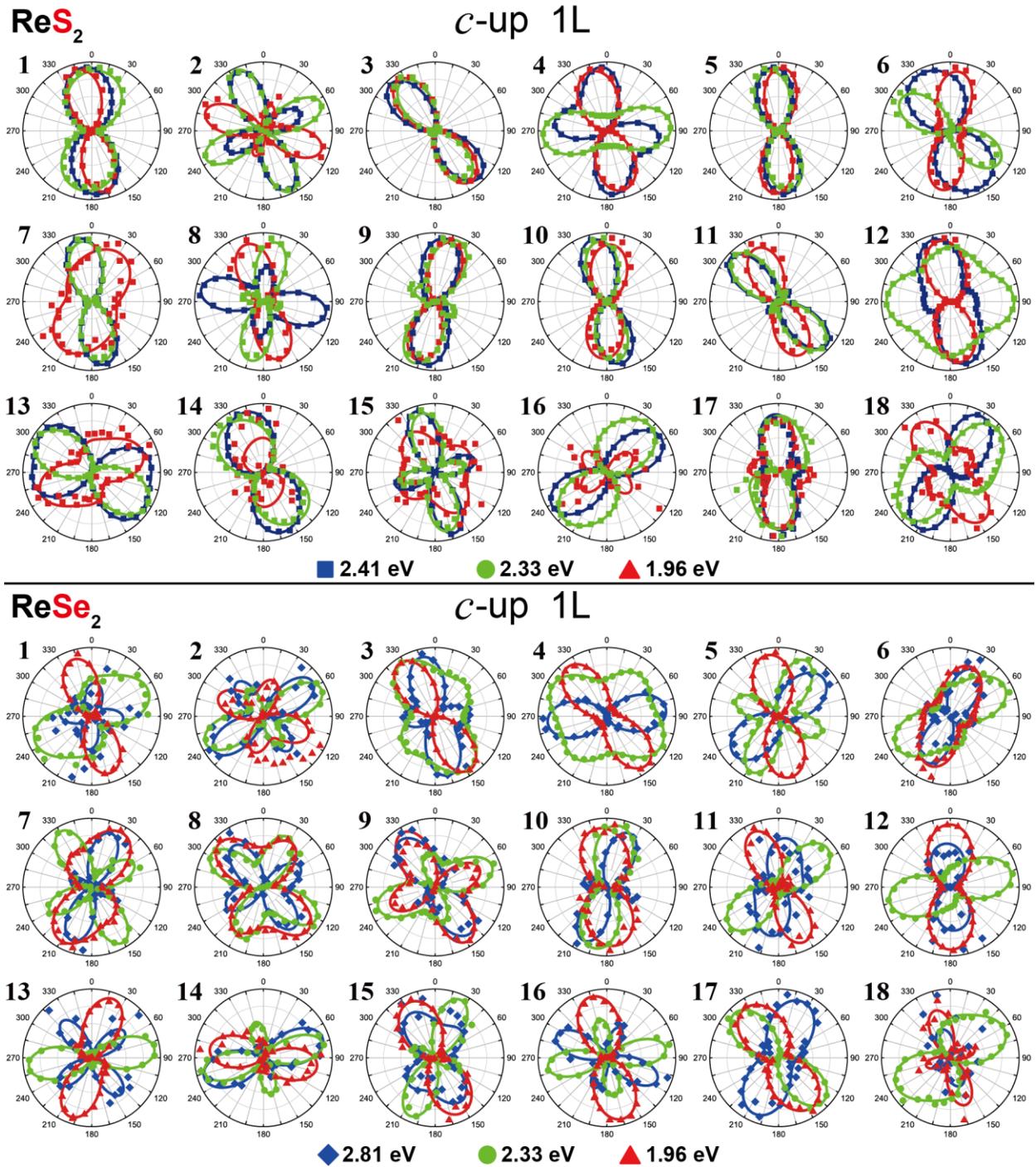

**Fig. S1** Polarization dependences of 18 Raman modes of '*c*-up' type monolayer ReS$_2$ and ReSe$_2$ measured with three different excitation energies as indicated.

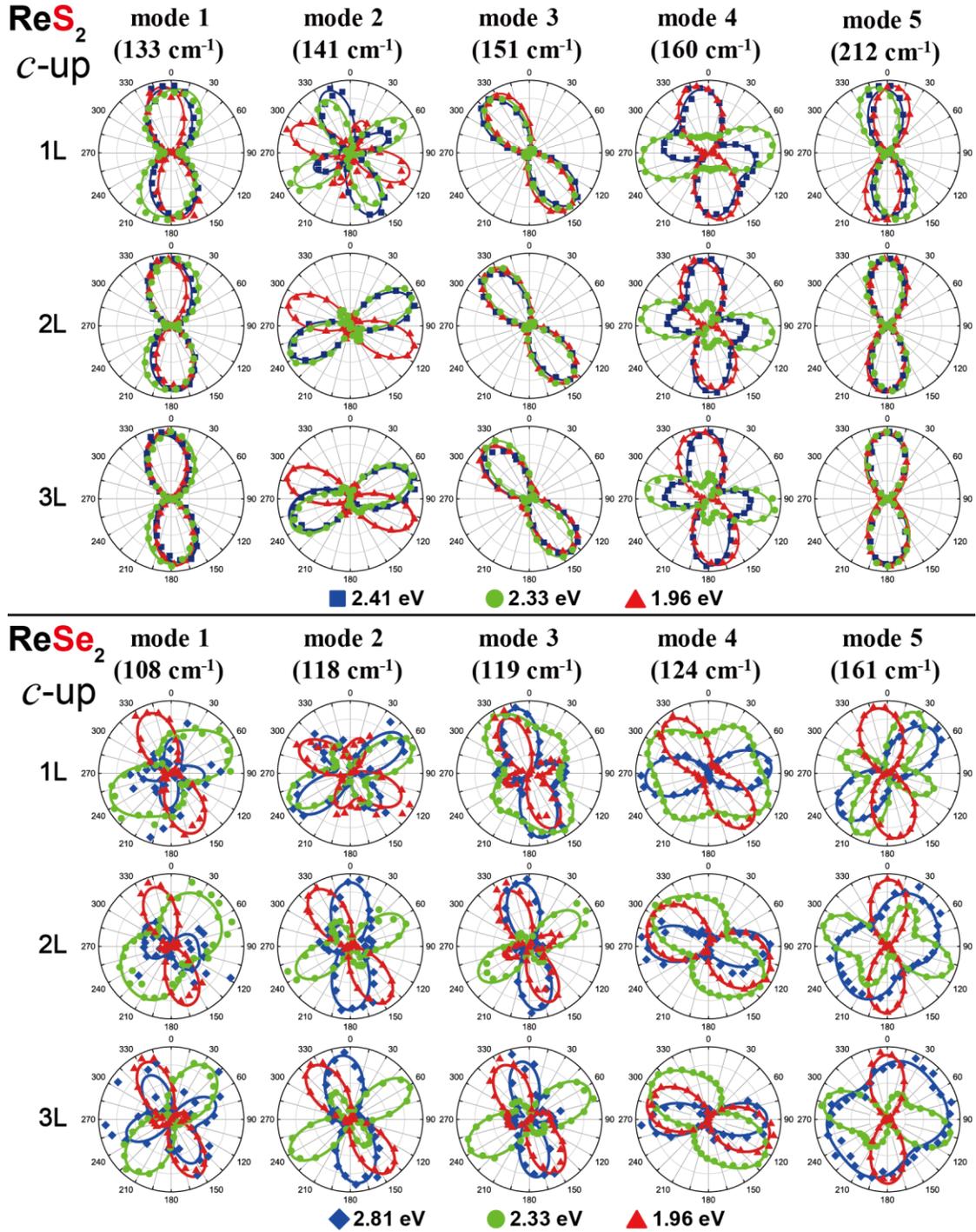

**Fig. S2** Polarization dependence of modes 1 to 5 of '*c*-up' type few-layer ReS$_2$ and ReSe$_2$.

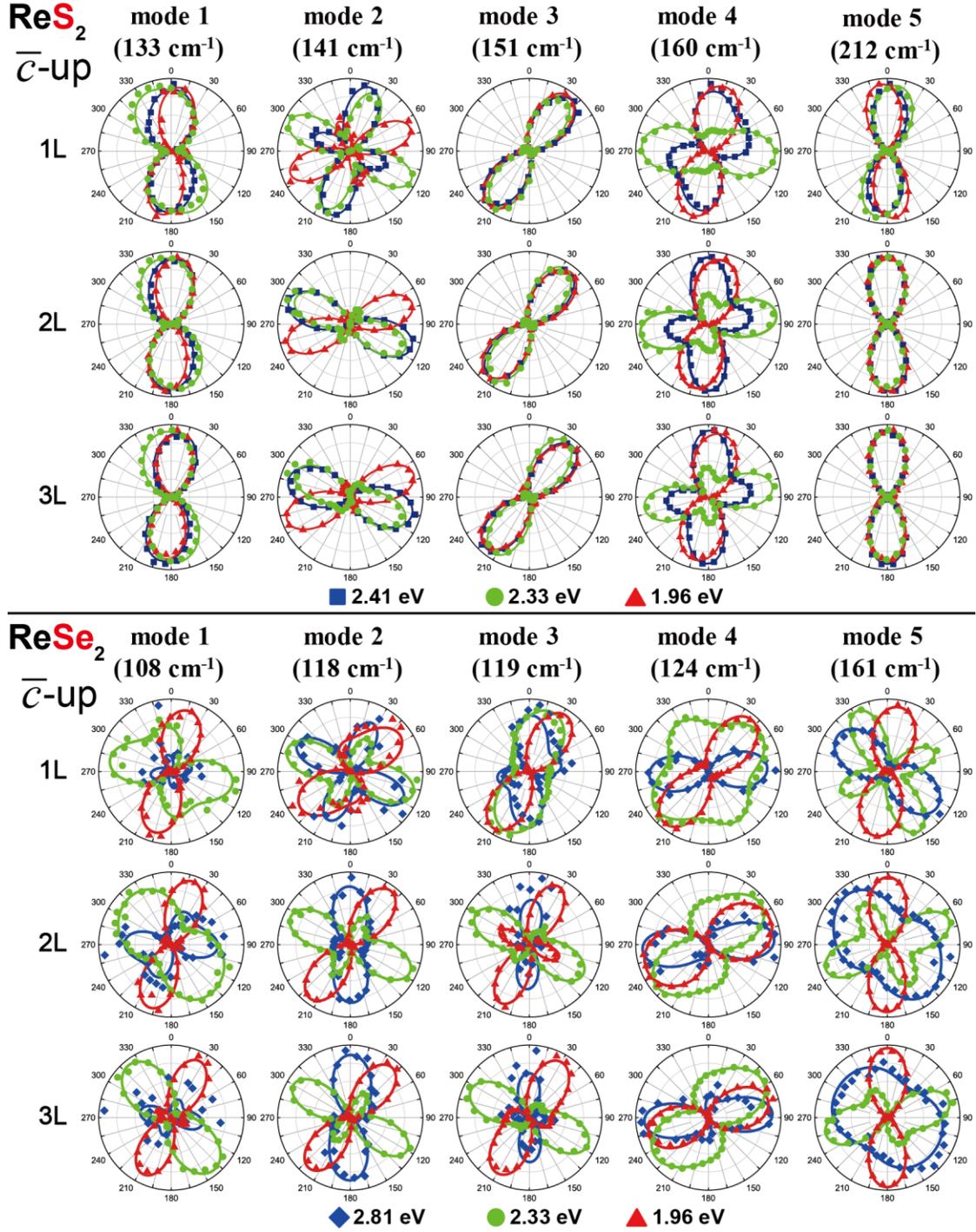

**Fig. S3** Polarization dependence of modes 1 to 5 of '$\bar{c}$-up' type few-layer ReS$_2$ and ReSe$_2$.

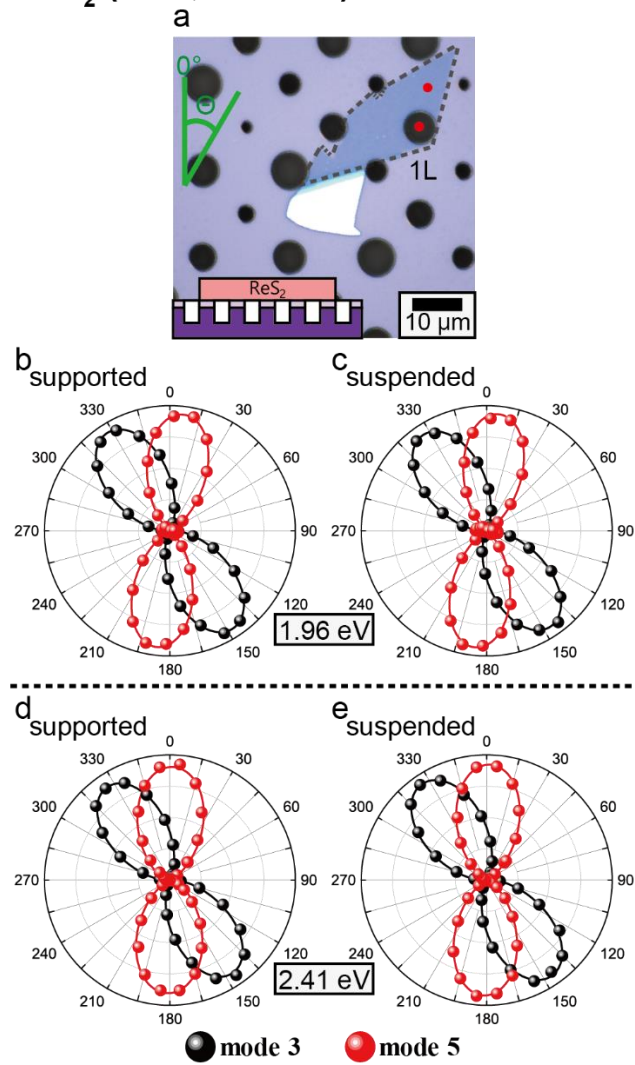

**Fig. S4** Polarization dependence of free-standing 1L ReS$_2$ sample. (a) Optical image of free-standing 1L ReS$_2$ sample. The measured spots are indicated by red dots. (b, c) Polarization dependences of 1L ReS$_2$ sample measured with the 1.96-eV laser. (d, e) Polarization dependences of 1L ReS$_2$ sample measured with the 2.41-eV laser. (b, d) sample on substrate (supported) (c, e) free-standing (supported) The polarization dependences of modes 3 and 5 show no difference between the supported and the suspended parts.

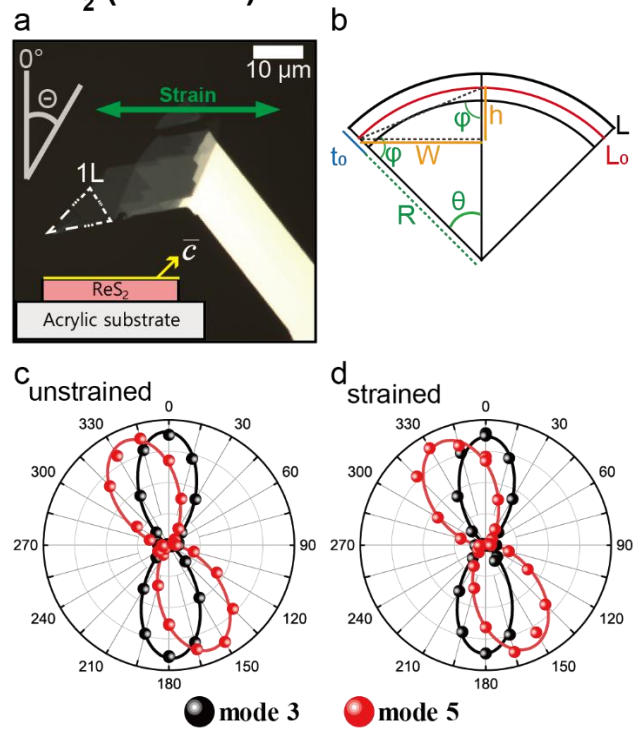

**Fig. S6** Effect of strain on polarization dependences of modes 3 and 5 of 1L ReS$_2$. (a) Optical image of the sample on acrylic substrate. (b) Schematic of strain estimation.[1] (c, d) polarization dependences of modes 3 and 5 for (c) unstrained and (d) 1.2%-strained sample. The uniaxial strain direction is taken as 0°.

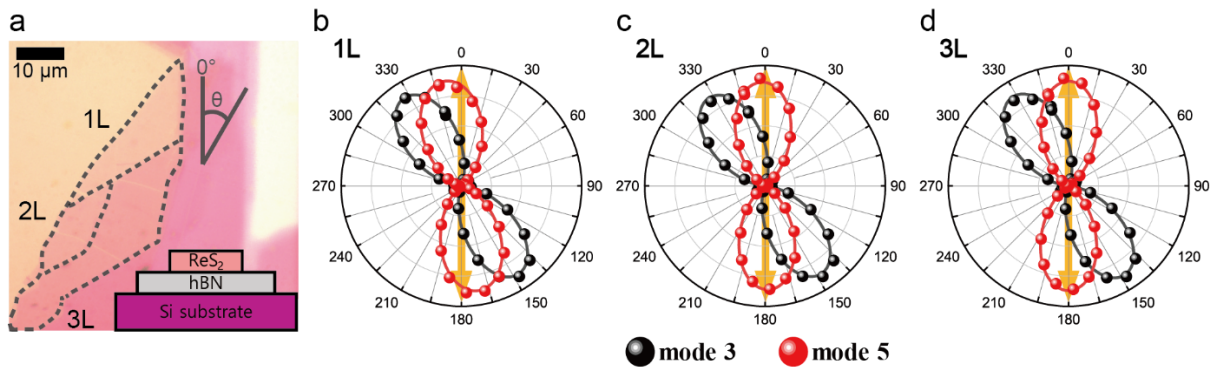

**Fig. S7** Polarization dependences of modes 3 and 5 from 1L to 3L ReS$_2$ on hBN

**References**

1    F P Beer, E R Johnston, J T Dewolf and D F Mazurek, *Mechanics of Materials*, McGraw Hill, 2012.